\documentclass[notitlepage]{article}  
\usepackage[utf8]{inputenc}
\usepackage{packages/packages}
\usepackage{graphicx}
\graphicspath{ {./} }
\usepackage[hidelinks]{hyperref}
\usepackage{physics, mathtools, xcolor}
\usepackage[shortlabels]{enumitem}
\usepackage{marginnote}
\usepackage{mdframed, hyperref}
\newcommand{\bC}{\mathbf{C}}
\newcommand{\bG}{\mathbf{G}}
\newcommand{\bO}{\mathbf{O}}
\newcommand{\bX}{\mathbf{X}}
\newcommand{\bk}{\mathbf{k}}
\newcommand{\bP}{\mathbf{P}}
\newcommand{\cT}{\mathcal{T}}
\newcommand{\cX}{\mathcal{X}}
\newcommand{\cI}{\mathcal{I}}
\newcommand{\link}[2]{\href{#1}{{\color{blue}#2}}}

\title{A Response to Economics as Gauge Theory}
\author{Timothy Nguyen\footnote{tim@timothynguyen.org}}
\date{}

\begin{document}

\maketitle

\begin{abstract}
    We provide an analysis of the recent work by Malaney-Weinstein on ``Economics as Gauge Theory" presented on November 10, 2021 at the Money and Banking Workshop hosted by University of Chicago. In particular, we distill the technical mathematics used in their work into a form more suitable to a wider audience. Furthermore, we resolve the conjectures posed by Malaney-Weinstein, revealing that they provide no discernible value for the calculation of index numbers or rates of inflation. Our conclusion is that the main contribution of the Malaney-Weinstein work is that it provides a striking example of how to obscure simple concepts through an uneconomical use of gauge theory. 
\end{abstract}

\section{Introduction}

Economic index numbers measure how prices or quantities have changed relative to some base period. This leads to the problem of formulating a reasonable notion of comparison, since the utility or even availability of the goods being compared varies across time. Such a problem was recognized and met with proposals for a solution at least as early as the work of Fisher and Shell\footnote{The author thanks David Baqaee for pointing out this reference.} \cite{Fisher1972TasteAQ}. 

In the recent work \cite{EG}, Malaney-Weinstein suggest that the right framework to handle time-dependent preferences is through the mathematical tools of gauge theory. This assertion goes back to the work of Malaney's thesis \cite{Malaney1996TheIN}, in which it shown that gauge theory offers a geometric solution to the defining and unifying of various index numbers. For our purposes, gauge theory can be described as the geometry of performing global comparisons\footnote{In physics, one compares internal states of particles (e.g. the spin of electrons).},
and so is suggestive as a toolkit for the economic problems at hand. Nevertheless, \cite{Malaney1996TheIN} has had little impact on the field of economics while \cite{EG}, being formulated using sophisticated mathematics, has yet to convince economists of its purported contribution.

The purpose of this work is to shed light on \cite{EG} based on the author's expertise in gauge theory.\footnote{The author however has no training in economics and so this paper should be read with that caveat in mind.} In particular, we answer the conjectures posed in \cite{EG} in a way that reveals that the contribution of \cite{EG} appears quite discouraging: namely the work veils easily understood concepts via a gratuitous use of gauge theory. Our findings can be summarized as follows:
\begin{itemize}
    \item Corollary 1 of \cite{EG}, when unpacked into a statement concerning economics, is a tautology.
    \item Conjecture 1 of \cite{EG} appears plainly false unless further clarified.
    \item Conjecture 2 of \cite{EG}, as a statement formulated in terms of gauge theory, is true for straightforward and unsurprising mathematical reasons. It is of no consequence to the computation of index numbers or inflation much less the field of economics at large.
\end{itemize}

Although the author has the mathematical training needed to appreciate some of the abstractions involved in \cite{EG}, given Weinstein's bold claim that economists ``don't know how to calculate inflation [because] they don't know gauge theory"\footnote{See this \link{https://youtu.be/XNS1Qs2n0Uc?t=6101}{interview}.}, the author felt obliged to call this assertion into question via the analysis presented herein. 

Our paper proceeds as follows. In Section 2, we review the work of \cite{EG} in a way that is intended to be more broadly accessible to a technical audience, namely only proficiency with calculus is expected in order to absorb most of the details. To facilitate this, we intersperse our exposition with a two-dimensional model example, which will appear as boxed text for ease of identification. Understanding the model example will enable the essential ideas of \cite{EG} to be gleaned, as working in higher dimensions only increases the abstraction without any additional technical innovation. In Section 3, we resolve the conjectures posed in \cite{EG}. Finally, in the appendix we provide additional insight into our resolution of the conjectures.\\

\noindent\textit{Note to reader:} It may be helpful for a first passing to read just the two-dimensional example in boxed text along with the appendix, referring to the general discussion only as needed.

\section{Setup}

We review the setup of \cite{EG} and simplify matters as needed. 

\subsection{Explication of \cite[Section 2]{EG}}

\begin{figure}[t!]
    \centering
    \includegraphics[scale=0.2]{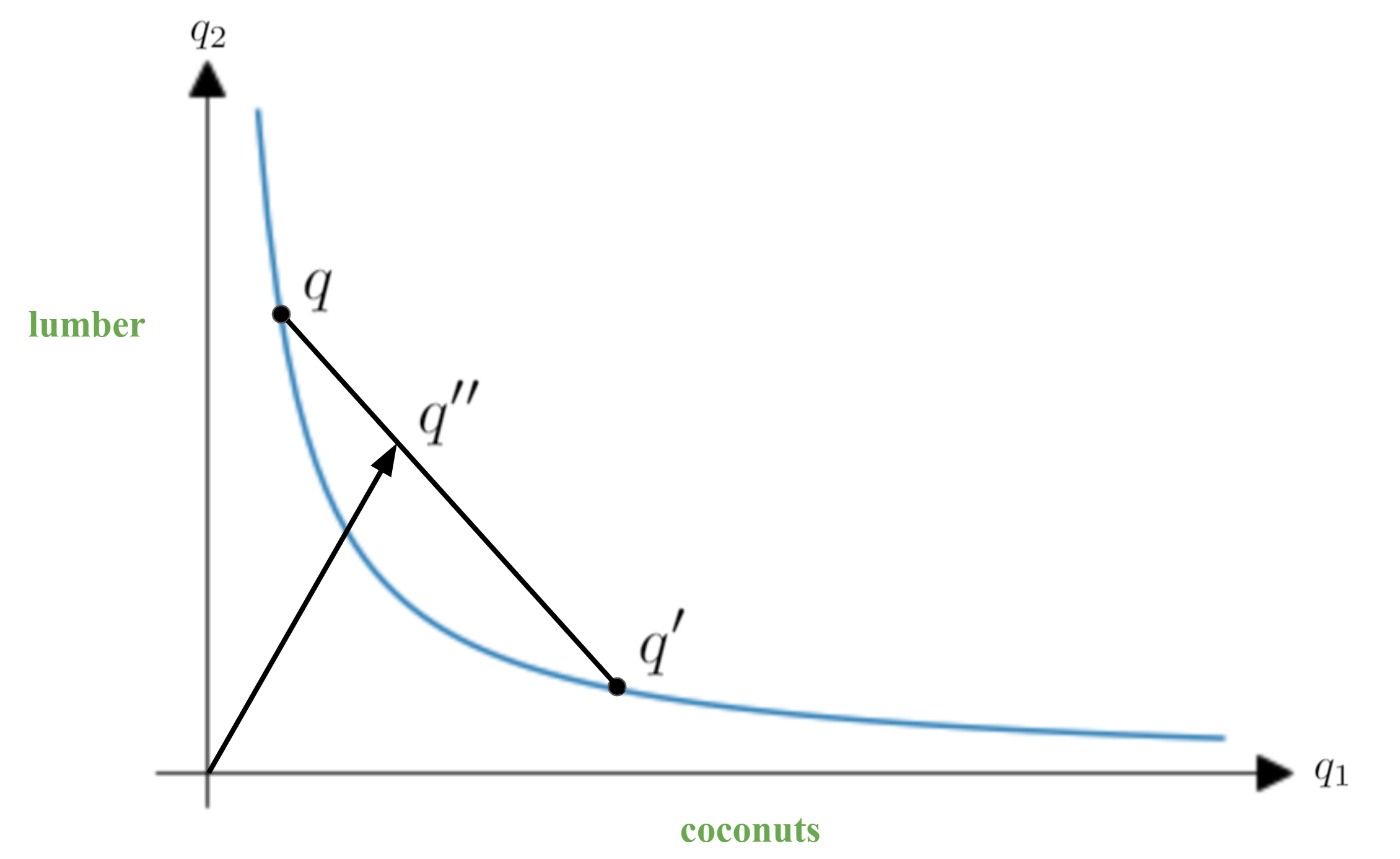}
    \caption{Illustration of convex to origin.}
    \label{fig:convex}
\end{figure}

Let $V = \R^n$ corresponding to an economy with $n$ goods. The positive orthant $V_+ = \{(q_1, \ldots, q_n) : q_1, \ldots, q_n > 0\}$ represents the subspace in which each good has positive quantity\footnote{It is a bit unusual that no good can occur in zero amount, but this assumption makes the \cite{EG} analysis more clean since then $V_+$ is a manifold (as opposed to a manifold with corners were we to instead consider the nonnegative orthant).}.
\begin{Definition}
A \textit{cardinal utility function} is a map
\begin{equation}
\bC: V_+ \to \R_+    
\end{equation}
such that for every $u \in \R_+$, we have $\bC^{-1}(u)$ is 
\begin{enumerate}
    \item  complete\footnote{Intuitively, this means the space cannot come to an abrupt end, which together with the convex to the origin condition implies that the space must have infinite extent in all directions (see Figure \ref{fig:convex}). However, this seems unnatural from a practical standpoint. It implies that the agent is completely averse to having zero of one of the goods: the agent would only agree to this if given an infinite amount of some other set of goods to compensate.} (as a metric subspace of $V_+$);
    \item convex to the origin, i.e. for any distinct $q$, $q' \in \bC^{-1}(u)$ and any point $q''$ lying on the open segment determined by $q$ and $q'$, the ray joining the origin to $q''$ intersects $\bC^{-1}(x)$ at some unique point prior to reaching $q''$. 
\end{enumerate}
To each cardinal utility function $\bC$, we obtain a \textit{foliation} of $V_+$ given by 
\begin{equation}
    \mathbf{O}_C = \{\bC^{-1}(u) : u \in \R_+\}.
\end{equation}
Each $\bC^{-1}(u)$ is a \textit{leaf} of the foliation and represents an indifference level set (the set of points with the same level of utility).  

We denote by $\bX_\bC: \R_+ \to V_+$ any \textit{cross-section}, i.e. 
a map that satisfies
\begin{equation}
    \bC(\bX_\bC(u)) = u, \qquad \textrm{for all }u \in \R_+.
\end{equation}
In other words, for each $u$, $\bX_\bC(u)$ is a particular basket of utility $u$.  
\end{Definition}
The commutative diagram \cite[(2.3)]{EG} is simply the statement that to evaluate a utility function $\bC$, one can either evaluate it directly, or else determine which foliation it belongs to (apply $\bO_C$) and then read off the cost from the level set value (the map $\mathbf{k}_\bC$).

\begin{mdframed}
\textbf{Model Example:} Let $n=2$ so that we have two goods (say coconuts and lumber for specificity as in \cite{Malaney1996TheIN}) in the amounts $q_1$ and $q_2$. Let 
\begin{equation}
    \bC(q_1,q_2) = q_1q_2, \label{eq:modelC}
\end{equation}
i.e. the utility of a basket of goods $(q_1,q_2)$ is the product of its two quantities. It is easy to verify that $\bC$ satisfies the properties needed to be a cardinal utility function. Indeed, the completeness property follows from the fact that the level sets 
\begin{equation}
\bC^{-1}(u) = \{(q_1, q_2) \in \R^2_+: q_1q_2 = u\} \label{eq:levelset}
\end{equation}
are all hyperbolas which are infinite in extent, while the convex to the origin property is easy to verify visually (see Figure \ref{fig:convex}). 

Our foliation $\mathbf{O}_C$ of $V_+$ is simply the carving out of $\R^2_+$ by the hyperbolas (\ref{eq:levelset}) as $u$ varies. Each such $u \in \R_+$ determines an \textit{indifference curve} since it consists of precisely those baskets that have utility $u$. 

We can choose the cross-section
\begin{equation}
    \bX_\bC(u) = (\sqrt{u}, \sqrt{u}) \label{eq:modelX}
\end{equation}
which picks the basket of equal weight between the two goods for every utility $u$. The image of $\bX_\bC$, denoted by $\mr{im}(\bX_\bC)$, consists of a unique representative basket from each indifference curve $\bC^{-1}(u)$. 
\end{mdframed}

\begin{figure}[t!]
    \centering
    \includegraphics[scale=0.2]{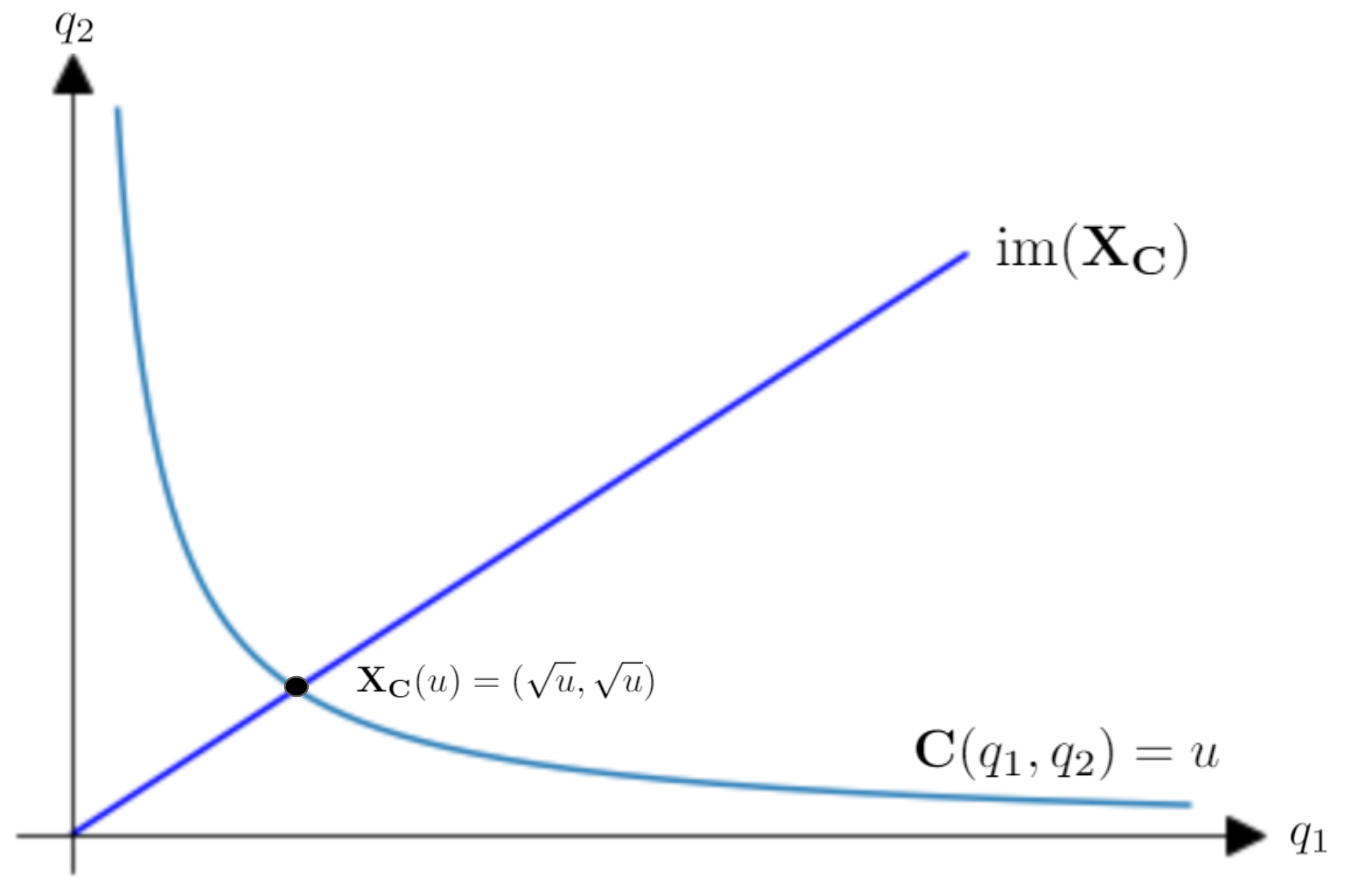}
    \caption{Model example where $\bC(q_1,q_2) = q_1q_2$ has level sets given by hyperbolas, and the image of the section map $\im \bX_\bC$ is a straight line given by (\ref{eq:modelX}) as $u$ varies.}
    \label{fig:imX}
\end{figure}

\subsection{Explication of \cite[Section 3]{EG}}

Given a cardinal utility function we obtain a foliation by indifference level sets, each level set sharing the same utility. On the other hand, given a (suitable) foliation, we can consider the family of all possible cardinal utility functions that give rise to the same foliation. We say such a cardinal utility function is \textit{compatible} with the foliation. Naturally, the only way in which two such utility functions can be compatible with a given foliation is that they differ only in the utility they assign to their shared collection of indifference level sets. That is, if $\bC_1$ and $\bC_2$ are two cardinal utility functions such that
\begin{equation}
    \bO_{\bC_1} = \bO_{\bC_2}
\end{equation}
then we must have $\bC_2 = \bG \circ \bC_1$ for some $\bG \in \mr{Diff}_+(\R_+)$, the space of orientation-preserving\footnote{The orientation-preserving condition is to ensure that ordinality of preferences are preserved, i.e. if $q$ is preferred over $q'$ because it has higher utility, then the same holds true between $\bG(q)$ and $\bG(q')$.} diffeomorphisms. That is, $\bG$ is a smooth monotonically increasing reassignment of utilities (with smooth universe). Any reassignment of utilities also gives us a new section simply by relabelling the utility of the basket assigned:
\begin{equation}
    \bX_\bC \mapsto \bX_\bC \circ \bG^{-1}.
\end{equation}

\begin{mdframed}
\noindent\textbf{Model Example:} Suppose
\begin{equation}
    \bG(u) = 2u,
\end{equation}
that is, $\bG$ doubles utility. If $\bC_1$ is given by (\ref{eq:modelC}) then define
\begin{equation}
    \bC_2 = (\bG \circ \bC_1)(q_1, q_2) = 2q_1q_2.
\end{equation}
The set indifference curves of $\bC_1$ and $\bC_2$ are the same: they consist of the hyperbolas $\{(q_1, q_2) : q_1q_2 = u\}$ as $u$ varies. However, if $\bC_1$ assigns utility $u$ to a hyperbola, then $\bC_2$ assigns it $2u$. 

If $\bX_1 = \bX_\bC(u) = (\sqrt{u}, \sqrt{u})$ is a utility $u$ representative basket for $\bC_1$ then
\begin{equation}
\bX_2 = (\bX_\bC \circ \bG^{-1})(u) = \bX_\bC(u/2) = (\sqrt{u/2}, \sqrt{u/2}) \label{eq:XCG-1}    
\end{equation}
is a utility $u$ representative basket for $\bC_2$. Moreover (\ref{eq:XCG-1}) shows that $\bX_1$ and $\bX_2$ pick out the same set of representative baskets (i.e. they have the same image) and differ only in the utility assigned to each such basket (as $u$ varies in (\ref{eq:XCG-1}), the corresponding basket has utility $u/2$ and $u$ with respect to $\bC_1$ and $\bC_2$, respectively).
\end{mdframed}

One can reformulate the above construction in terms of bundle language if one wants to adopt the perspective of a differential geometer. Thus, there is a total space $\cT = \cT_{\cC\cX}$ of all possible cardinal utility functions and cardinal cross sections $(\bC, \bX_\bC)$. There is a projection map $\Pi^{\cT}$ which coarsifies the $\bC$ to its set of indifference level sets and replaces $\bX_\bC$ by its induced map on the corresponding foliation:
\begin{equation}
    \Pi^{\cT}(\bC, \bX_\bC) = (\bO_\bC, \bX_\bC \circ \bk_\bC) \label{eq:Pi}
\end{equation}
The preimage of (\ref{eq:Pi}) under $\Pi^{\cT}$ is the set
\begin{equation}
    \{(\bG \circ \bC, \bX_\bC \circ \bG^{-1}): \bG \in \mr{Diff}_+(\R_+)\} \label{eq:fiber}
\end{equation}
i.e., the set of all utility reparameterizations. Thus, $\mr{Diff}_+(\R_+)$ serves as the ``gauge-group" acting on $\cT$, since it parameterizes invariances with respect to the projection $\Pi^{\cT}$.

The infinite-dimensional nature of the problem captures two aspects of the current setup: (1) all possible baskets are treated simultaneously rather than a single basket (2) all possible cardinal utility functions (and hence indifference level sets) are considered rather than a single one. This is why the construction becomes more involved, albeit not in any conceptually mysterious way. 

\subsection{Explication of \cite[Section 4]{EG}}

Suppose we now have a price function $\bP: V_+ \to \R$ which we take to be a linear and positive function. Given a foliation of indifference level sets $\bO$, pick any cardinal utility function $\bC$ that is compatible with $\bO$, i.e. such that $\bO = \bO_\bC$ (in what follows, the construction is independent of the particular choice of $\bC$). Our hypotheses imply that each indifference level set has a unique basket whose price is minimal along the level set.\footnote{Completeness implies existence, convex to the origin implies uniqueness. We omit the details of a proof.} This allows us to define a distinguished section\footnote{Our notation here slightly differs from \cite{EG}, which is more involved since \cite{EG} works on a quotient space whereas we work with an explicit representative $\bO = \bO_\bC$. Since what follows is independent of the representative $\bC$, there is no loss in using our less verbose notation.}
\begin{align}
    \bX^{\bP}_\bC(u) &= \{\textrm{the unique point on }\bC^{-1}(u)\textrm{ that minimizes } \bP\} \\
    &= \mr{argmin}_{q \in V_+} \{\bP(q): \bC(q) = u\}
    \label{eq:XPO}
\end{align}
which picks out this \textit{minimal price basket} along the indifference level set with utility $u$.

The previous construction allows us to define a function $m$ which assigns to any pair $(\bO, \bP)$ a corresponding pair of a cardinal utility function and section:
\begin{equation}
    m(\bO, \bP) = (\bP \circ \bX^\bP_\bC \circ \bC, \bX^\bP_\bC) \in \T. \label{eq:m}
\end{equation}
In words, the right-hand side consists of the following. The section $\bX^\bP_\bC$ assigns to a utility $u \in \R_+$ the minimal price basket along the indifference level set of $\bO = \bO_\bC$ given by $\bC^{-1}(u)$. The cardinal utility function $\bP \circ \bX^\bP_\bC \circ \bC: V_+ \to \R_+$ assigns to $q \in V_+$ the value determined by a two-step process: (i) computing the utility of $q$ under $\bC$; (ii) returning the price of the minimal price basket of utility equal to that obtained in (i) (by applying $\bP \circ \bX^\bP_\bC$).

The above was adapted from \cite{EG}. We find it helpful to reformulate the definition of $m$ into a form that may be more recognizable when we compute the cost-of-living adjustment (COLA) or Konus index, following \cite{diewert2004economic}: 
\begin{Definition}\label{def:cost} (Reformulation)
Define the \textit{cost-of-living function}
\begin{equation}
    C_\bC(u, \bP) = \min_{q \in V_+} \{\bP(q): \bC(q) = u\},
\end{equation}
measuring the cost to acquire utility $u$ for a price-minimizing agent given the price level $\bP$.
We have that (\ref{eq:m}) can be written as
\begin{equation}
    m(\bO, \bP) = \left(C_\bC(\cdot, \bP), \bX^\bP_\bC\right) \label{eq:m2}
\end{equation}
where we have chosen any $\bC$ compatible with $\bO$.
\end{Definition}

Following \cite{diewert2004economic}, we can define the COLA index under dynamic utility functions and prices $(\bC_t, \bP_t)$ at time $t$ relative to some basic period $t_0$ via
\begin{equation}
    \mathcal{I} = \frac{C_{\bC_t}(u_t, \bP_t)}{C_{\bC_{t_0}}(u_{t_0}, \bP_{t_0})},
\end{equation}
so long as we can decide on how to choose the reference utilities $u_t$ and $u_0$. This will be the purpose of the subsequent sections, though we will ultimately use the expression $\bP \circ \bX^\bP_\bC \circ \bC$ for computations involving the cost-of-living function.\\

\begin{mdframed}\textbf{Model Example:}
Consider $\bP(q_1,q_2) = q_1 + q_2$, i.e., the price of our basket of two goods is equal to the sum of the individual quantities. Then it is easy to check that $\bX^\bP_\bC$ is given by $\bX_\bC$ in (\ref{eq:modelX}). Indeed, this follows from the fact that $\bX^\bP_\bC(u)$ must be the point along the indifference curve $\bC^{-1}(u) = \{(q_1, q_2): q_2 = u/q_1\}$ whose tangent line has slope $-1$ (since then the gradient of $\bP$ will be perpendicular to such a line, as is required by the method of Lagrange multipliers when finding optima).

Then
\begin{equation}
    m(\bO_\bC, \bP) = (\tilde \bC, \bX_\bC)
\end{equation}
where
\begin{align}
    \tilde \bC(q_1,q_2) &= \bP \circ \bX^\bP_\bC \circ \bC(q_1,q_2)\\
    &= \bP \circ \bX_\bC(q_1q_2)\\
    &= \bP(\sqrt{q_1q_2}, \sqrt{q_1q_2}) \\
    &= 2\sqrt{q_1q_2}
\end{align}

\end{mdframed}

If we have a one-parameter family of pairs of ordinal preferences and prices \begin{equation}
\alpha(t) = \left(\bO(t), \bP(t)\right),\qquad  t \in [t_a, t_b]
\end{equation}
we get an associated path $\tilde \alpha = (m \circ \alpha)(t)$ in $\cT$.\\

\noindent\textbf{Notation.} Given a function of time $f(t)$, we may write $f_t$ instead of $f(t)$.\\

\subsection{Explication of \cite[Sections 5--6]{EG}}

This section is purely to complete our exposition of \cite{EG}, though it is optional and not essential for our response. First time readers can skip ahead to the next section.

Section 6 of \cite{EG} constructs a particular connection on the infinite-dimensional space $\cT$, which we will call the \textit{Malaney-Weinstein connection}. It has a similar flavor to the connection constructed in Malaney's PhD thesis \cite{Malaney1996TheIN}, in that it identifies particular subspaces related to the price function in order to define an associated covariant derivative. The Malaney-Weinstein connection however is infinite-dimensional while the connection in \cite{Malaney1996TheIN} is finite-dimensional. 

We summarize (in a simplified manner) the Malaney-Weinstein connection in a few words and then return to our two-dimensional case for an explicit illustration. A connection tells us which variations of a $(\bC, \bX_\bC) \in \cT$, i.e. which tangent vectors are to be considered ``horizontal". A tangent vector to $(\bC, \bX_\bC)$ has two components $\gamma: V_+ \to \R$ and $\nu: \mr{im}(\bX_\bC) \to \R^n$ corresponding to the variation in $\bC$ and $\bX_\bC$ respectively (essentially $\gamma$ is the function $\frac{d}{dt}\big|_{t=0}\bC_t$ where $\bC_0 = \bC$ and $\nu$ is a vector field along the image of $\bX_\bC$). Then $(\gamma, \nu)$ is horizontal if and only if
\begin{enumerate}
    \item $\gamma$ vanishes along $\mr{im}(\bX_\bC)$; 
    \item for every $q \in \mr{im}(\bX_\bC)$, we have $\nu(q)$ belongs to the tangent space to the indifference level set of $\bC$ passing through $q$. 
\end{enumerate}
Intuitively, the above conditions state that the tangent vector $(\gamma, \nu)$ is horizontal if to first order, the resulting perturbation does not affect minimal prices and moves minimal price baskets only along their associated indifference curves. 

\begin{mdframed}
\textbf{Model Example:} We have $\bC$ and $\bX_\bC$ as given by (\ref{eq:modelC}) and (\ref{eq:modelX}). Then $(\gamma, \nu)$ is parallel with respect to the Malaney-Weinstein connection if and only if for every $u \in \R_+$
\begin{enumerate}
\item $\gamma: V_+ \to \R$ satisfies $\gamma(\sqrt{u}, \sqrt{u}) = 0$ for all $u$;
\item $\nu: \mr{im}(\bX_\bC) \to \R^n$ is a scalar multiple of the vector $(1, -1)$ at every point.
\end{enumerate}

\end{mdframed}

\subsection{Explication of \cite[Sections 7--9]{EG}}
\label{sec:7-8}

The previous section introduced a connection, the bread and butter of gauge theory, on the infinite-dimensional space $\cT$. As such, an appreciable amount of geometric abstraction was involved. The reason why the details are not crucial is that the significance of a connection is through its associated \textit{parallel transport} operators, which can easily be understood in terms of ordinary differential equations.

If we abstract the situation, it becomes easier to understand this parallel transport operator. Suppose we have a curve $\tilde\alpha: [t_a, t_b] \to G$, where $G$ is some connected Lie group (for simplicity take it to be $\R$ under addition)\footnote{Technically, $\tilde\alpha$ is valued in what is known as a principal $G$-bundle rather than $G$, but we overlook this for simplicity.}. A connection induces a notion of what it means for a section to be \textit{parallel}. When our connection is a particular distinguished connection, the \textit{trivial connection}, the notion of parallel coincides with that of constant:
\begin{center}
\begin{tabular}{cccc}
\begin{tabular}{c}
    $\tilde\alpha(t)$ is parallel\\ (using the trivial connection) 
    \end{tabular} & $\Leftrightarrow$ & $\displaystyle \frac{d}{dt}\tilde\alpha(t) = 0$, & for all $t \in [t_a, t_b].$\\
    \end{tabular}
    \end{center}
But for a general connection, we have to adjust $\tilde\alpha$ in order for it to be parallel. Namely, we find a uniquely determined group-valued map $\Upsilon^{\tilde\alpha}_{t_a}: [t_a, t_b] \to G$
such that\footnote{In \cite{EG}, $\Upsilon^{\tilde\alpha}$ is described in terms of right multiplication whereas we use left multiplication in (\ref{eq:upsilon-def}). This reflects the difference between working with sections of an adjoint bundle globally (which requires using the global right multiplication on a principal bundle) versus working in a local trivialization in our simplified presentation, in which left-multiplication is available. Non gauge-theorists need not be concerned about this technical subtlety as it does not affect the analysis that follows.}
\begin{equation}
    t \mapsto \Upsilon^{\tilde\alpha}_{t_a}(t)\cdot\tilde\alpha(t)  \textrm{ is parallel}, \qquad \Upsilon^{\tilde\alpha}_{t_a}(t_a) = \mr{id}. \label{eq:upsilon-def}
\end{equation}
In other words, $\Upsilon^{\tilde\alpha}_{t_a}$ adjusts $\tilde\alpha$, a curve starting at $t_a$, to being parallel via pointwise multiplication. Writing $\Upsilon^{\alpha}_{t,t'} = \Upsilon^{\tilde\alpha}_{t'}(t)$, the following properties are obeyed (which expresses that the $\Upsilon^{\tilde\alpha}$ form a \textit{$1$-parameter group of transformations}):
\begin{alignat}{2}
\Upsilon^{\tilde\alpha}_{t,t} &= \mr{id} \qquad & \textrm{(identity)} \label{eq:Upsilon1}\\
\Upsilon^{\tilde\alpha}_{t_b,t_a} & = \left(\Upsilon^{\tilde\alpha}_{t_a,t_b}\right)^{-1} & \textrm{(invertibility)} \label{eq:Upsilon2}\\
\Upsilon^{\tilde\alpha}_{t_c,t_b}\Upsilon^{\tilde\alpha}_{t_b,t_a} &= \Upsilon^{\tilde\alpha}_{t_c,t_a} & \textrm{(compositionality)}  \label{eq:Upsilon3}
\end{alignat}

In \cite{EG}, the group $G$ corresponds to the group $\mr{Diff}_+(\R_+)$, the gauge group of $\cT$. Thus, for each $t$, we can think of $\Upsilon^{\alpha}_{t,t_a}$ in this setting as a utility or price adjustment diffeomorphism, such that if $\tilde\alpha: [t_a, t_b] \to \T$, then the path $t \mapsto  \Upsilon^{\tilde\alpha}_{t,t_a} \cdot \tilde\alpha_t$ is parallel. 

Let $\cO$ denote the space of all ordinal preferences (given by foliations arising from cardinal utility functions) and $\cP$ denote the space of all price functions. Then since 
\begin{equation}
    m: \cO \times \cP \to \cT
\end{equation}
as defined by (\ref{eq:m}), $m$ defines a bundle over $\cO \times \cP$ whose fiber over $(\bO, \bP)$ is the fiber in $\cT$ containing $m(\bO, \bP)$. Denote this bundle $m^*(\cT)$. 

Moreover, $m$ also defines a section of $m^*(\cT)$. We also write $\tilde \alpha$ to denote this section. This defines for us a trivial connection on $m^*(\cT)$, the one that treats $\tilde \alpha$ as constant\footnote{Note that a trivial connection depends on a trivialization of a bundle, or equivalently, a choice of section.}. We also have the induced Malaney-Weinstein connection on $m^*(\cT)$ given by pulling back the one on $\cT$.

\begin{figure}[t!]
    \centering
    \includegraphics[scale=0.2]{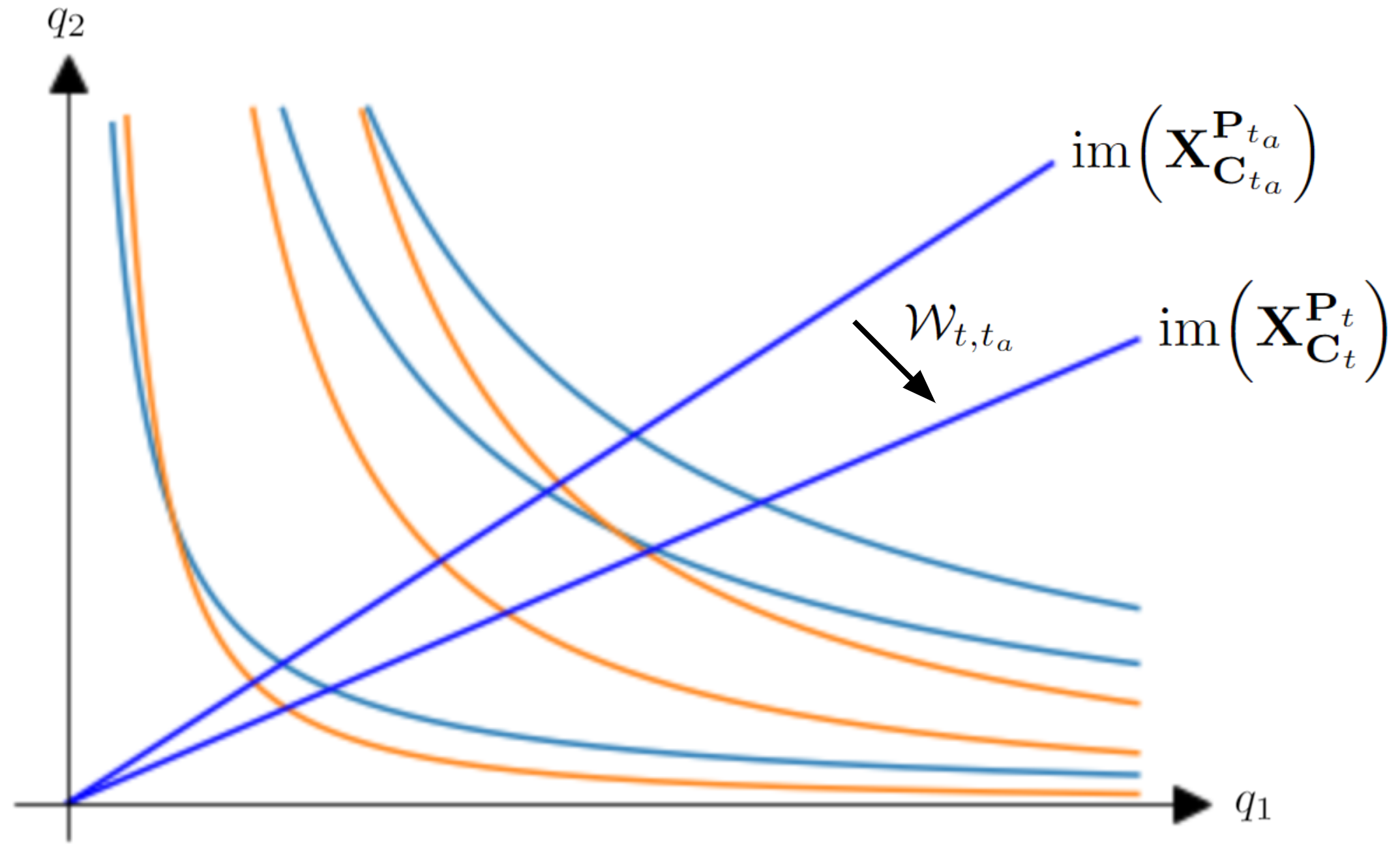}
    \caption{Welfare maps are determined by what they do on minimal cost baskets: $\W_{t,t_a}$ maps the baskets $\im\!\!\left(\bX^{\bP_{t_a}}_{\bC_{t_a}}\right)$ to the baskets in $\im\!\!\left(\bX^{\bP_t}_{\bC_t}\right)$. Light blue curves are indifference level sets of $\bC_{t_a}$, orange curves are indifference are level sets of $\bC_t$.}
    \label{fig:welfare}
\end{figure}

\begin{mdframed}
\textbf{Model Example:} Suppose we have the family of cardinal utility functions
\begin{equation}
    \bC_t(q_1, q_2) = q_1^tq_2, \qquad t \in [1,2].
\end{equation}
Let $\bP_t(q_1, q_2) = q_1 + q_2$ be constant in $t$. Thus, in words, the contribution to the utility of the first commodity (coconuts) varies as $q_1^t$ while prices stay fixed.

So the $\bO_t = \bO_{\bC_t}$ are each a foliation of $V_+$ given by the level sets of $\bC_t$. The corresponding minimal price basket function $\bX_{\bC_t}^{\bP_t}$ for a utility $u$ can be found from the method of Lagrange multipliers by solving the system of equations
\begin{equation}
\begin{rcases}
\begin{aligned}
    \pd_{q_1}F_t(q_1, q_2) &= 0\,\\
    \pd_{q_2}F_t(q_1,q_2) & = 0\\
    q_1^tq_2 &= u
\end{aligned}
\end{rcases}
\end{equation}
where $F_t(q_1, q_2) = q_1 + q_2 + \lambda (q_1^tq_2 - u)$. A simple computation shows that the minimizer, as a function of $u$, is given by $q_t^* = (q_{1,t}^*, q_{2,t}^*)$ where
\begin{align}
    q_{1,t}^*(u) &= t^{\frac{1}{t+1}}u^{\frac{1}{t+1}} \label{eq:q1*u}\\
    q_{2,t}^*(u) &= t^{-\frac{t}{t+1}}u^{\frac{1}{t+1}}. \label{eq:q2*u}
\end{align}
It is more convenient to express this minimizer in terms of cost rather than utility. We have via Definition \ref{def:cost}
\begin{align}
    C_{\bC_t}(u, \bP_t) &= \bP_t(q^*_t(u)) \label{eq:cost*}\\
    &= q_{1,t}^*(u) + q_{2,t}^*(u) \\
    &= \left(t^{\frac{1}{t+1}} + t^{-\frac{t}{t+1}}\right)u^{\frac{1}{t+1}}. \label{eq:c(u)}
\end{align}
Hence, denoting the left-hand side of (\ref{eq:cost*}) by $c$, by abuse of notation, we can use (\ref{eq:c(u)}) to rewrite (\ref{eq:q1*u}) and (\ref{eq:q2*u}) in terms of $c$:
\begin{align}
    q_{1,t}^*(c) &= \frac{tc}{1+t} \label{eq:q1*c}\\
    q_{2,t}^*(c) &= \frac{c}{1+t} \label{eq:q2*c}.
\end{align}
Note that by definition of $c$, we have
\begin{equation}
    \bP_t(q_t^*(c)) = c.
\end{equation}

\noindent\textbf{Question: } What is a sensible way to compute the price index at time $t$ relative to the (prior) base time $t_a = 1$ given the evolving preferences $\bO_t$ and prices $\bP_t$?\\ 

Answering this question involves two parts: deciding which reference basket of goods to consider for the base time and then deciding upon a fair comparison of goods at a future time. For the first of these, while the computation of a single index number uses a single reference basket, the approach considered here is to consider all possible baskets (as parameterized by the minimal baskets in $\im( \bX^{\bP_t}_{\bC_t})$), each of which will give rise to their own (not necessarily the same) price index. For the second part, we have the notion of a \textit{welfare map} (see Figure \ref{fig:welfare})
\begin{equation}
    \cW_{t,t_a}: \bO_{t_a} \to \bO_t
\end{equation}
mapping indifference curves at time $t_a$ to those at time $t$ (two curves related by $\cW$ are understood to represent equal amounts of welfare). Because an indifference curve can be specified by its minimal price basket, a welfare map can also be expressed as a correspondence between such minimal price baskets:
\begin{equation}
    \cW_{t,t_a}: \im\!\!\left(\bX^{\bP_{t_a}}_{\bC_{t_a}}\right) \to \im\!\!\left(\bX^{\bP_t}_{\bC_t}\right).
\end{equation}
In our present example, the set of minimal price baskets at any given time is especially simple. Based on (\ref{eq:q1*c})--(\ref{eq:q2*c}), as the cost $c$ varies, the minimal baskets at time $t$ sweeps out the line
\begin{equation}
    \mr{im} (\bX_{\bC_t}) = \{(q_1, q_2) : q_2 = q_1/t\}.
\end{equation}

In the (unnecessary) bundle setting described above, welfare maps are determined by a choice of parallel transport operator, which can either be the one given by \cite[Section 7]{EG}, the naive or default ``parallel equals constant" parallel transport operator, or else any other parallel transport operator in principle. But we can study these welfare maps directly without such sophisticated machinery as we will now do.

Consider the naive welfare maps $\cW^{\mr{naive}}_{t,t_a}$ given by the naive parallel transport operator. It says that we should treat costs as constant in time, in which case we have $C' = \cW^{\mr{naive}}_{t,t_a}(C)$ if and only if the indifference curves $C'$ and $C$ have the same cost with respect to $\bX^{\bP_t}_{\bC_t}$ and $\bX^{\bP_{t_a}}_{\bC_{t_a}}$ respectively. Indeed, this is exactly expressing the parallel condition with respect to the trivial connection. Rephrasing this in terms of minimal price baskets, we have 
\begin{equation}
    \cW^{\mr{naive}}_{t,t_a}(q^*_{t_a}(c)) = q^*_{t}(c). 
\end{equation}

Next, consider the welfare map $\cW^{MW}_{t,t_a}$ proposed by Malaney-Weinstein. Then
\begin{equation}
    \cW^{MW}_{t,t_a}(q^*_{t_a}(c)) = q^{MW}_{t}(c),
\end{equation}
for some $q^{MW}_{t}(c)$ lying on the line $q_2 = t^{-1}q_1$. We have
\begin{equation}
q^{MW}_t(c) = q^*_t(\Upsilon_{t,t_a}^{\tilde\alpha}(c))  
\end{equation}
because by definition $\Upsilon_{t,t_a}^{\tilde\alpha}(c)$ is the cost adjustment needed to induce parallelism with respect to the Malaney-Weinstein connection. 

The resulting COLA price indices can be computed as follows. For the naive welfare map, we have the ratio of prices
\begin{align}
    \cI^{\mr{naive}} 
    &= \frac{\bP_t(q^*_{t}(c))}{\bP_{t_a}(q^*_{t_a}(c))}\\
    &= \frac{c}{c} \\
    & = 1
\end{align}
for every $c$. This is expected of our naive welfare map: if prices do not change and we naively assume that constant welfare across times equates to constant cost, then the price index stays fixed at $1$. 

For the Malaney-Weinstein welfare map, we have 
\begin{align}
    \cI^{MW} &= \frac{\bP_t(q^{MW}_t(c))}{\bP_{t_a}(q^*_{t_a}(c))} \label{eq:IMW1}\\
    &= \frac{\bP_t\left(q^{*}_t(\Upsilon_{t,t_a}^{\tilde\alpha}(c))\right)}{\bP_{t_a}(q^*_{t_a}(c))} \label{eq:IMW2}\\
    &= \frac{\Upsilon_{t,t_a}^{\tilde\alpha}(c)}{c} \label{eq:IMW3}
\end{align}
precisely recovering Corollary 1 of \cite{EG}. Note however there is nothing distinguished about the Malaney-Weinstein connection in deriving (\ref{eq:IMW3}). Every connection gives a corresponding parallel transport operator, i.e. cost adjustment operator, and so the derivation \textit{(\ref{eq:IMW1})--(\ref{eq:IMW3}) holds for any connection}, where $\Upsilon_{t,t_a}^{\tilde\alpha}$ is the corresponding cost-adjustment operator of that particular connection. 

In plain English therefore, Corollary 1 of \cite{EG} asserts the following:
\begin{equation}
\begin{tabular}{|c|}\hline
    \textit{The most general cost of living adjustment index is given by}\\ \textit{ the ratio of an arbitrarily chosen adjusted price to a base price.}
\\ \hline
\end{tabular}\label{eq:Cor}
\end{equation}

\vspace{.1in}

There is nothing surprising about that statement, except possibly by obscuring it through a choice of price adjustment that invokes infinite dimensional gauge theory. While our above analysis of Corollary 1 was done for our model example, it extends straightforwardly to the general case. 
\end{mdframed}

\section{Resolution of Conjectures of \cite{EG}}

We resolve the conjectures of \cite{EG} in the below Proposition. While we will continue to refer to our model example to help illustrate the proof, it is still rather involved since the model example still requires an infinite-dimensional analysis (working in two-dimensions does not eliminate the fact that the space of all possible utility functions is infinite-dimensional). Thus, we provide a finite-dimensional analog of our analysis in the appendix (the gauge group is finite-dimensional instead of infinite-dimensional) so that it may illuminate our analysis here. First time readers may want to start there before returning to the main proof.

\begin{Proposition}\label{mainprop}
We have the following:
\begin{enumerate}
    \item Conjecture 2 of \cite{EG} is true and can be restated without gauge theory in a way that makes the result simple. See (\ref{eq:Conj2}).
    \item Conjecture 1 of \cite{EG} appears false unless further clarified.
\end{enumerate}
\end{Proposition}

\Proof (i) Conjecture 2 is essentially just the following assertion:
\begin{equation}
\begin{tabular}{|c|}\hline
    \textit{All possible ways of tracking time-dependent preferences and prices}\\ \textit{is equivalent to all possible ways of tracking minimal price baskets.}
\\ \hline
\end{tabular}\label{eq:Conj2}
\end{equation}
(Of course, the above is restricted to the setting in which our technical hypotheses on cardinal utility functions and prices hold, so that the notion of minimal price baskets is well-defined.) Indeed, if we look at equation (9.3) of \cite{EG}, the equation says that a welfare map that takes indifference level sets at time $t_b$ to those at time $t_a$ is given by
\begin{enumerate}[1.]
\item computing time $t_b$ costs ($\cE_{t_b}$, which in our notation is the cost-of-living function $C_{\C_{t_b}}$)
\item applying cost adjustments ($\Upsilon_{t_b,t_a}^{\tilde\alpha}$)\item finding the corresponding indifference level sets at time $t_a$ ($\cE_{t_a}^{-1}$).
\end{enumerate}
The content of Conjecture 2 is to ask what is the structure of all possible choices of $\Upsilon^{\tilde\alpha}$, not just the one arising from Conjecture 1 via the particular choice of the Malaney-Weinstein connection. We saw a related phenomenon in the Model Example of Section \ref{sec:7-8}, where at the conclusion of our analysis, we saw how the COLA price index could be made arbitrary because we are free to choose any cost adjustment $\Upsilon^{\tilde\alpha}$ satisfying (\ref{eq:Upsilon1})--(\ref{eq:Upsilon3}). In the setting of Conjecture 2, we have a similar situation: we can either define $\Upsilon^{\tilde\alpha}$ arbitrarily and get a correspondence between indifference level sets, or else we can start with any correspondence between indifference level sets and then $\Upsilon^{\tilde\alpha}$ is defined accordingly.  
This freedom is precisely the freedom of choosing an adjoint-bundle valued $1$-form as we now explain.

By the general setup of gauge theory, parallel transport operators are equivalent to the choice of a covariant derivative. This is because infinitesimal parallel transport determines a covariant derivative and conversely, solving the relevant ordinary differential equations arising from a covariant derivative yields parallel transport operators. A $1$-parameter family of welfare operators is the same as a $1$-parameter group of parallel transport operators. Considering all such $1$-parameter group of operators leads us to consider all possible parallel transport operators, i.e. all covariant derivatives. The space of all covariant derivatives is an affine space over one-forms with values in the adjoint bundle of the gauge group \cite{kobayashi:1963}. Thus, connecting all the dots, we can summarize the equivalences just described as follows:

\begin{align}
\begin{tabular}{|ccccc|} \hline
    \hspace{-.1in}\begin{tabular}{c}
    parallel transport operators \\ (welfare operators)\end{tabular}\hspace{-.1in} & $\Leftrightarrow$ & \hspace{-.1in}\begin{tabular}{c}
    covariant \\ derivatives\end{tabular}\hspace{-.1in} & $\Leftrightarrow$ & \hspace{-.1in}\begin{tabular}{c}$1$-forms valued in\\
    the adjoint-bundle
    \end{tabular}\hspace{-.1in} \\ \hline 
\end{tabular} \label{eq:equivalences}
\end{align}

Hence, because of (\ref{eq:equivalences}), Conjecture 2 holds true via standard gauge theory. The plain English rephrase (\ref{eq:Conj2}) follows from noting that parallel transport is determined by how it acts on sections (and thus how it acts on the minimal price baskets given by the sections $\bX^\bP_\bC$). Thus the space of all welfare operators (which is meant to track utility under time-dependent preferences and prices) is determined by how it operates on minimal price baskets (how it parallel transports them, which in turn, is determined by the data of an adjoint-bundle valued $1$-form). Altogether, the rendition of Conjecture 2 into (\ref{eq:Conj2}), once all technical distractors from \cite{EG} are stripped, reveals its remarkable simplicity.

\begin{mdframed}
\textbf{Model Example: } A general welfare map is such that
\begin{equation}
    \cW_{t,t_a}(q^*_{t_a}(c)) = q^*_t(f_{t,t_a}(c))
\end{equation}
for some $f_{t,t_a} \in G = \mr{Diff}_+(\R_+)$. In words, $f_{t,t_a}$ maps the minimal price basket of cost $c$ at time $t_a$ to the minimal price basket of cost $f_{t,t_a}(c)$ at time $t$. For $\W_{t,t_a} = \W^{\mr{naive}}_{t,t_a}$ equal to the naive welfare map, $f_{t,t_a}(c) = c$ for all $t$, and for $\W_{t,t_a} = \W^{MW}_{t,t_a}$ associated to the Malaney-Weinstein connection, $f_{t,t_a} = \Upsilon_{t,t_a}^{\tilde\alpha}$. But in general, we can choose general $f_{t,t_a}$ that forms a $1$-parameter group of diffeomorphisms.
\end{mdframed}

(ii) Conjecture 1 appears to be false because the naive welfare map is a valid welfare map satisfying the hypotheses of Conjecture 1. (Note that the properties stated in Conjecture 1 are exactly reflected in the $1$-parameter group properties (\ref{eq:Upsilon1})--(\ref{eq:Upsilon3})).
The naive welfare map is formed out of the trivial connection defined with respect to the section $\tilde\alpha = m$ and which underlies the construction of the elements occurring in Conjecture 1, see \cite[(8.3)]{EG}. We believe what Malaney-Weinstein intended Conjecture 1 to state is that there is some canonical way to arrive at the Malaney-Weinstein connection akin to how the Levi-Civita connection from Riemannian geometry naturally arises as a canonical connection.\footnote{The Levi-Civita connection is the unique connection on the tangent bundle of a Riemannian manifold which is both metric and torsion-free.} It is not obvious to the author in what sense the Malaney-Weinstein connection may be canonical. Nevertheless, even if Conjecture 1 were amended so as to be true (which would require clarifying what Malaney-Weinstein means by ``constructible"), it possesses no clear significance without further elaboration.\End

\section{Conclusion}

In this paper, we resolve the conjectures of Malaney-Weinstein by disentangling the gauge theory from the economics. While the application of gauge-theory to economics could potentially be fruitful (and the author would be very enthusiastic about such a positive outcome), the particular attempt of \cite{EG} in its current state falls far short of what could be called a success. Namely, the complexity that went into the constructions of \cite{EG} far exceeded the gains of what was produced:
\begin{itemize}
    \item Corollary 1 is a tautology as can be verified directly from our reformulation (\ref{eq:Cor}).
    \item Conjecture 1 asserts that a certain way of relating dynamic preferences is unique in some sense, but that sense is not sufficiently defined and so renders Conjecture 1 false. Even if Conjecture 1 were to become true through an adequate clarification of the ``constructible" criterion, further work would be required to establish any theoretical or practical significance.
    \item Conjecture 2 is a statement about all possible ways of relating dynamic preferences. It is automatically true as posed from a gauge-theoretic perspective. It too has no theoretical or computational consequences. In fact, because it is a statement about all possible dynamic preferences, as seen by (\ref{eq:Conj2}), it is a statement about all possible ways of obtaining a price index. It thus offers no insights into how to calculate particular price indices. Thus, from a practical standpoint, Conjecture 2 is vacuous. 
\end{itemize}
We hope that the technical analysis we have provided here enables those who wish to verify for themselves our conclusions about \cite{EG}.

\bibliographystyle{alpha}
\bibliography{references}

\begin{thebibliography}{MW21}

\bibitem[Die]{diewert2004economic}
E~Diewert.
\newblock The economic approach to index number theory: the single-household
  case.
\newblock {\em Consumer Price Index Manual--Theory and Practice, Gen{\`e}ve},
  pages 313--335.
\newblock
  \url{https://www.ilo.org/public/english/bureau/stat/download/cpi/ch17.pdf}.

\bibitem[FS72]{Fisher1972TasteAQ}
Franklin Fisher and Karl Shell.
\newblock {\em Taste and Quality Change in the Pure Theory of the True
  Cost-of-Living Index}, pages 1--48.
\newblock 12 1972.

\bibitem[KN63]{kobayashi:1963}
S.~Kobayashi and K.~Nomizu.
\newblock {\em {Foundations of Differential Geometry}}, volume~1.
\newblock Wiley Classics, 1963.

\bibitem[Mal96]{Malaney1996TheIN}
Pia Malaney.
\newblock The index number problem: A differential geometric approach.
\newblock 1996.
\newblock Ph.D. Thesis. Chapter 2 coauthored with Eric Weinstein.

\bibitem[MW21]{EG}
Pia~N. Malaney and Eric~R. Weinstein.
\newblock {An Extention of Intertemporal Ordinal Welfare to Changing Tastes:
  Economics as Gauge Theory}.
\newblock
  \url{https://economics.uchicago.edu/sites/economics.uchicago.edu/files/Welfare_Chicago_Draft.pdf},
  2021.
\newblock Upload Date: Novebmer 10, 2020.

\end{thebibliography}

\appendix 

\section{Simplified explication of Conjecture 2}

We provide a finite-dimensional analogue of the analysis that went into the proof of Proposition \ref{mainprop}. We want to illustrate the correspondences (\ref{eq:equivalences}) and how that reveals the tautology underlying Conjecture 2. Suppose we are working in $n$-dimensions, where we take $n = 1$ for maximal simplicity. Let our gauge group be $G = \R$, which means that the objects we will be considering (the ``sections of $G$-bundles") will be real-valued functions.

Let $p, q \in \R$ be two points. We want to relate copies of $G$ at $p$ and $q$. We can do this by parallel transporting $G_p$ to $G_q$ (the copies of $G$ at $p$ and $q$, respectively). A parallel transport map has to be $G$-\textit{equivariant}, which in our case, simply means that our parallel transport operator\footnote{What we call $\tau$ here is almost nearly identical to what is called $\Upsilon^{\tilde \alpha}$ in the main text. The latter measures the affect of parallel transport on $\tilde\alpha$.} $\tau$ is determined by what it does on a unique point - the rest is determined by translation. That is, if
\begin{align}
    \tau_{q,p}: G_p & \to G_q
\end{align}
sends $x \in G_p$ to $\tau_{q,p}(x)$, then  \begin{equation}
\tau_{q,p}(x + x') = \tau_{q,p}(x) + x' \qquad \textrm{for all } x' \in G. 
\end{equation}
The determination of $\tau_{q,p}(x)$, for any given $x$, is obtained by solving a differential equation. Namely, let\footnote{Without loss of generality, let $q \geq p$.} $\gamma: [t_0, t_1] \to \R$ be a the straight line joining $p$ to $q$, i.e.
\begin{equation}
\gamma(t) = p + t, \qquad t_0 = 0 \leq t \leq t_1 = q-p.    
\end{equation}
We solve the initial value problem for the function $f$
\begin{equation}
    \begin{rcases}
\begin{aligned}
    \frac{d}{dt} f(t) + A(\gamma(t))f(t) &= 0\\
    f(0) &= x 
\end{aligned}
\end{rcases}\label{eq:1d-diffeq}
\end{equation}
and then set
\begin{equation}
    \tau_{q,p}(x) = f(t_1). \label{eq:1d-upsilon}
\end{equation}
Here $A$ is a function which we refer to as a \textit{connection}. The operator $\frac{d}{dt} + A$ occurring in (\ref{eq:1d-diffeq}) is the covariant derivative associated to $A$. In general, $A$ is an adjoint-bundle valued $1$-form, but for the present $1$-dimensional case with $G = \R$, we can identity such $1$-forms with functions\footnote{The adjoint condition is vacuous in the abelian setting, in particular, for $G = \R$.}. The equivalences (\ref{eq:equivalences}) can be restated in the present simplified one-dimensional case as
\begin{align}
\begin{tabular}{|ccccc|} \hline
    (\ref{eq:1d-upsilon}) & $\Leftrightarrow$ & (\ref{eq:1d-diffeq}) & $\Leftrightarrow$ & $A$ \\ \hline 
\end{tabular} \label{eq:1d-equivalences}
\end{align}
That is, the collection of parallel transport operators $\tau_{q,p}$ (as $p$ and $q$ vary) is in one-to-one correspondence with the differential equations (\ref{eq:1d-diffeq}) which in turn is in one-to-one correspondence with the space of all possible $A$. This is manifest: $A$ determines a differential equation which determines a unique solution and this chain of logic can be reversed. Thus, we have established (\ref{eq:1d-equivalences}) in the 1-dimensional case, but it holds more generally in arbitrary dimension and for arbitrary gauge group.

Conjecture 2 of \cite{EG} is just the assertion (\ref{eq:equivalences}), which is  the infinite-dimensional analogue of (\ref{eq:1d-equivalences}). Namely, Conjecture 2 posits that the space of all welfare maps is given by the space of all possible objects  playing the role of $A$ as above. Based on the above, Conjecture 2 holds true from the general tenets of gauge theory. While (\ref{eq:equivalences}) is a foundational part of the gauge theory setup, it is a well-understood piece of mathematics with no demonstratable significance to economics. Indeed, the way it is used in \cite{EG} appears to be merely to assert (\ref{eq:Conj2}).
\end{document}